\documentclass[3p,final,10pt]{elsarticle}

\usepackage[T1]{fontenc}
\usepackage[latin9]{inputenc}
\usepackage{amsmath}
\usepackage{amssymb}
\usepackage{esint}
\usepackage{microtype}
\usepackage{subfig}
\usepackage{booktabs} 
\usepackage{amsfonts}
\usepackage{arydshln} 
\usepackage{mathtools}
\usepackage{lineno}   
\usepackage{mathrsfs}
\usepackage{MnSymbol} 
\usepackage{enumitem} 
\usepackage{pifont}
\usepackage{stmaryrd} 
\usepackage{wasysym}
\usepackage{marvosym}
\usepackage{tcolorbox} 
\usepackage{xcolor} 
\usepackage{hyperref}
\hypersetup{colorlinks=true, linkcolor=blue, citecolor=blue, urlcolor=blue}
\usepackage{xurl}
\makeatletter
\def\ps@pprintTitle{%
  \let\@oddhead\@empty
  \let\@evenhead\@empty
}
\newcommand{\Bpart}{\mathcal{B}^{p}}

\newcommand{\Sigmaijp}{\overline{\sigma}_{ij}^{\,p}}
\begin{document}
%
\begin{tcolorbox}[
    colback=gray!5, 
    colframe=gray!70, 
    arc=0mm, 
    width=\textwidth,
    boxrule=0.5pt,
    top=2mm, bottom=2mm, left=2mm, right=2mm
]
\small
\textbf{Please cite this article as:} \\
  Matthew R. Kuhn (2025),
  ``On average stress within sub-regions of granular media,''
  \emph{Mechanics Research Communications},
  Vol. 150,
  104567,
  https://doi.org/10.1016/j.mechrescom.2025.104567.
\end{tcolorbox}
\vspace{1em} 
\begin{frontmatter}
\title{On average stress within sub-regions of granular media}
\author[up]{Matthew~R.~Kuhn}
\address[up]{Donald P. Shiley School of Engrg., Univ. of Portland,
             5000 N. Willamette Blvd., Portland, OR, USA 97231}
%
%
%
\begin{abstract}
The paper concerns computation of average stress within
small sub-regions of a larger static granular assembly,
where the sub-region's boundary is allowed
to pass through the assembly's particles.
An exact average is computed for certain stress components
and for certain categories of sub-regions.
The paper also identifies those choices of sub-regions for which
exact stress components may be exactly computed,
but when these conditions are
not met, provides reasonable bounds on the error.
\end{abstract}
\begin{keyword}
  Granular material \sep stress \sep multi-scale methods
\end{keyword}
\end{frontmatter}
%
%
\section{\normalsize Introduction} \label{sec:introduction}
Multi-scale methods in mechanics rely on 
credible determinations of stress and strain at
multiple scales.
For granular media, these scales include the
molecular (smaller than the grains),
the micro-scale
(in which a grain is the fundamental unit),
the meso-scale (encompassing sub-assemblies of grains),
the macro-scale of a large assembly,
and the continuum scale,
in which a material's discrete nature is altogether abrogated.
Because measures of strain are
based on particle and point movements,
well-founded methods are available for evaluating stress
at the micro- and macro-scales.
Measuring stress in granular materials at the intermediate,
meso-scale is more problematic, and rational means for
measuring stress within small sub-regions
is the current focus.
\par
The most relevant definitions of stress are based upon
the inter-particle contact forces among particles,
and early approaches include
the works of
Love \cite{Love:1927a}, Weber \cite{Weber:1966a},
and Rothenburg \cite{Rothenburg:1981a}.
These works are clarified or extended by
considering a region's material-cell partition
\cite{Bagi:1996a,Satake:2004a},
by applying virtual work to grain systems
\cite{Bardet:2001a,Chang:2005a},
by considering couple stress
\cite{Bardet:2001a,Kruyt:2003a,Froiio:2006a},
by including body forces \cite{Bagi:1999a},
and by including inertia effects for dynamic systems
\cite{Nicot:2013b,Yan:2019a}.
Taken together,
these works provide means of determining the \emph{average stress}
for a granular assembly from the
contact forces among particles.
A stress derived in this manner can be used
to develop macro-level constitutive models from particle-scale
analyses or simulations.
However, to study localized phenomena within a larger assembly
(for example, in local regions having large gradients of
deformation or rotation),
one must focus on material behavior at meso-scales,
smaller than the entire assembly but larger than the
grains themselves~---
behavior in sub-regions
that encompass perhaps a few hundred particles that
are embedded within the entire
assembly \cite{Gaspar:2001a,Gaspar:2002a,Kuhn:2002a}.
While the works mentioned above can be applied
to sub-regions whose boundaries only pass through
contact points between particles,
more general sub-regions will have boundaries that
pass through the particles themselves,
and the contributions
of such \emph{peripheral
particles} cannot be ignored.
\par
The paper derives expressions (sometimes, approximations)
for average stress in granular sub-regions, accounting
for a boundary that passes through the sub-region's peripheral particles.
The average stress will be computed from two sets of
forces: internal
contact forces between particles and
external body forces acting on the grains.
Such body forces,
which include the usual presence of gravity,
can also be applied as an experimental means
to coerce non-uniform deformation (e.g., \cite{Kuhn:2002a}).
We disallow, however, body couples
and contact couples,
and we assume static equilibrium among the particles.
\par
We find that an exact average stress can not, in general, be determined
when the boundary passes through a sub-region's peripheral particles.
This imprecision arises from an uncertainty of the spatial distribution
of stress within these peripheral particles.
The paper, however, suggests special cases that eliminate
the imprecision, and it derives
bounds for the error in other cases.
%
\section{\normalsize Derivations}\label{sec:derivations}
\begin{figure}
\centering
\includegraphics[scale=1.00]{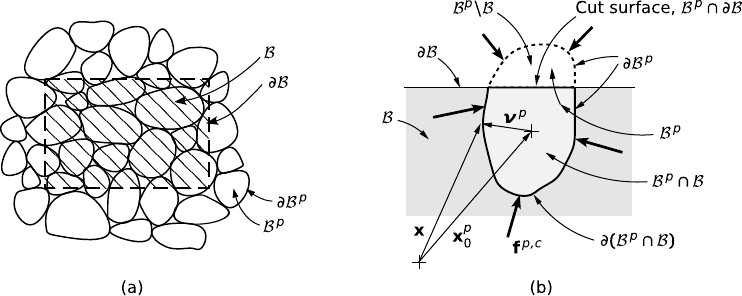}%
\caption{A faceted sub-region $\mathcal{B}$: (a)~location within a 
         larger region;
         (b)~a particle $\Bpart$
         intersected by sub-region $\mathcal{B}$,
         with contact forces $\mathbf{f}^{p,c}$ acting on $\Bpart$.}
\label{fig:stress-calc-all}
\end{figure}

We seek the average stress $\overline{\sigma}_{ij}^{\,\mathcal{B}}$
within a sub-region $\mathcal{B}$ that is embedded in a larger region:
\begin{equation} \label{eq:stress_def}
\overline{\sigma}_{ij}^{\,\mathcal{B}} =
\frac{1}{V^{\mathcal{B}}}
\underset{\mathcal{B}}{\int}
{\sigma_{ij}(\mathbf{x})\,dv}
\end{equation}
where $V^{\mathcal{B}}$ 
is the sub-region's volume,
and $\partial\mathcal{B}$ is its boundary
(Fig.~\ref{fig:stress-calc-all}a).
The region of a single, $p$-th, particle is $\Bpart$,
noting that the bodies
$\Bpart$ of \emph{peripheral particles}, those which straddle
$\partial\mathcal{B}$, will include material both inside and outside of
$\mathcal{B}$ (Fig.~\ref{fig:stress-calc-all}b).
%
%
The body $\Bpart$ is partitioned
as follows:
material within both $\Bpart$ and $\mathcal{B}$,
designated as $\Bpart \cap \mathcal{B}$;
material within $\Bpart$ that lies outside of
$\mathcal{B}$, designated as $\Bpart \backslash \mathcal{B}$;
and the \emph{cut surface}
of boundary $\partial\mathcal{B}$ that is shared with
$\Bpart$, written as
$\Bpart \cap \partial\mathcal{B}$
(Figs.~\ref{fig:stress-calc-all}b and~\ref{fig:stress-calc4b}).
For interior, non-peripheral particles,
$\Bpart \cap \mathcal{B} = \Bpart$ and
$\Bpart \backslash \mathcal{B} = \Bpart \cap 
 \partial\mathcal{B} = \varnothing$.%
\par
We restrict sub-region $\mathcal{B}$
to having a faceted, piecewise planar boundary $\partial\mathcal{B}$
in which all \emph{cut surfaces} are planar and are oriented 
in one of the coordinate directions, 
$\mathbf{e}_{1}$, $\mathbf{e}_{2}$, or $\mathbf{e}_{3}$
(e.g., a
rectangular box region $\mathcal{B}$ with edges aligned with the
coordinate axes).
This restriction will aid in isolating uncertainties
in the nine components of
$\overline{\sigma}_{ij}^{\mathcal{B}}$.
%
\par
We begin with a result of
Bagi \cite{Bagi:1999a} for
the average stress of a static
granular medium with body forces (volume forces), but in which
boundary $\partial\mathcal{B}$ passes only along or outside
the particles' surfaces:
\begin{equation} \label{eq:Bagi}
\overline{\sigma}_{ij}^{\,\mathcal{B}} \;=\; \frac{1}{V^{\mathcal{B}}}
\sum_{p\in\mathcal{B}}{\left(
\underset{\partial(\Bpart\cap\mathcal{B})}{\int}
\negthickspace\negthickspace\negthickspace
{x_{i}t_{j}^{p}\,ds} \;+
\underset{\Bpart\cap\mathcal{B}}{\int}
\negthickspace\negthickspace
{x_{i}b_{j}^{\,p}\,dv}
\right)}
\end{equation}
where
$\mathbf{b}^{p}$ is the vector field of body force density
within~$p$, and $\mathbf{t}^{p}$ is the vector field
of traction acting on $p$'s surface.
This result also applies when $\partial\mathcal{B}$ passes
through peripheral particles, with the understanding that
traction $\mathbf{t}^{p}$ 
is comprised of forces on two parts of a particle's surface.
On particles wholly within $\mathcal{B}$,
tractions are in the form of
idealized contact forces $\mathbf{f}^{p,c}$
(a contact $c$ between $p$ and its neighbors).
For peripheral particles,
traction $\mathbf{t}^{p}$ also includes
traction produced by the stress distribution
along the cut surface
$\Bpart \cap \partial\mathcal{B}$.
\par
With the particles in equilibrium, the total
external force on each $p$-th particle is zero:
\begin{equation} \label{eq:Q}
\left(
\underset{\Bpart\cap\partial\mathcal{B}}{\int}
\negthickspace
t_{j}^{p}\,ds \;+
\underset{\partial\Bpart\cap\mathcal{B}}{\int}
\negthickspace
t_{j}^{p}\,ds
\right)
\;+
\underset{\Bpart\cap\mathcal{B}}{\int}
\negthickspace
b_{j}^{\,p}\,dv = 0 , \quad j\in\{ 1,2,3\},\; p\in\mathcal{B}
\end{equation}
or
\begin{subequations}\label{eq:Q2}
\begin{gather}
\underset{\Bpart\cap\partial\mathcal{B}}{\int}
\negthickspace
t_{j}^{p}\,ds \;+
T_{j}^{p}
= 0 , \quad j\in\{ 1,2,3\},\; p\in\mathcal{B}\\
T_{j}^{p} \equiv
\underset{c\in (\partial\Bpart \cap \mathcal{B})}{\sum}
\negthickspace\negthickspace
{f_{j}^{\,p,c}}
\;+
\underset{\Bpart \backslash \mathcal{B}}{\int}
\negthickspace\negthickspace
{b_{j}^{\,p}\, dv}
\end{gather}
\end{subequations}
where the second integral in Eq.~(\ref{eq:Q}) is balanced by
the sum of those contact forces
$f_{j}^{\,p,c}$ acting wholly within sub-region $\mathcal{B}$,
not including its boundary $\partial\mathcal{B}$
(i.e., the sum in Eq.~\ref{eq:Q2}b).
This sum is included in the total
resultant external force $\mathbf{T}^{p}$,
a force essential in derivations below.
Traction along the cut surface $\mathcal{B}^{p}\cap\partial\mathcal{B}$
(the first integral in Eqs.~\ref{eq:Q} and~\ref{eq:Q2}a)
supplies the counteracting force to $\mathbf{T}^{p}$,
thus maintaining $p$'s equilibrium.
\par
Coordinates $x_{i}$ in Eq.~(\ref{eq:Bagi})
are referenced to a global, common
Cartesian frame.
More useful is a set of local, offset coordinates
$\nu_{i}^{\,p}$, local to each~($p$-th) particle and
are aligned with the global coordinate directions:
\begin{equation} \label{eq:local}
x_{i} = x_{0i}^{p} + \nu_{i}^{\,p}
\end{equation}
where $x_{0i}^{p}$ are the global coordinates of a reference material point attached to~$p$ (Fig.~\ref{fig:stress-calc-all}b).
Because each particle is in equilibrium,
all terms in
Eq.~(\ref{eq:Q2}) can be multiplied by
a particle's $x_{0i}^{p}$,
and then the collection of these equations for all particles
$p\in\mathcal{B}$
are subtracted from the summation in Eq.~(\ref{eq:Bagi})
to yield the following
(see~\cite{Bagi:1999a}):
\begin{subequations} \label{eq:Bagi2}
\begin{align}
\overline{\sigma}_{ij}^{\,\mathcal{B}}
&=
\frac{1}{V^{\mathcal{B}}}
\sum_{p\in\mathcal{B}}
V^{p}\, \Sigmaijp
\\
\Sigmaijp
&=
\frac{1}{V^{p}}
\left(
\underset{\Bpart\cap\partial\mathcal{B}}{\int}
\negthickspace\negthickspace
{\nu_{i}^{\,p}t_{j}(\boldsymbol{\nu}^{p})\,ds}
\;+
\sum_{c\in (\partial\Bpart\cap\mathcal{B})}
\negthickspace\negthickspace
{\nu_{i}^{\,p\,,c}f_{j}^{p,\,c}}
\;+
\underset{\Bpart\cap\mathcal{B}}{\int}
\negthickspace\negthickspace
{\nu_{i}^{\,p} b_{j}^{\,p} \,dv}
\right)
\end{align}
\end{subequations}
where $V^{p}$ is the volume of $p$.
In this equation, the global coordinates $x_{i}$ of
Eq.~(\ref{eq:Bagi})
are replaced by the local coordinates $\nu_{i}^{\,p}$,
showing that
the average stress $\overline{\sigma}_{ij}^{\,\mathcal{B}}$
is independent of the choice
of local offsets $x_{0i}^{p}$.
By considering the moment equilibrium of particles,
Bagi \cite{Bagi:1999a} demonstrated that
$\overline{\sigma}_{ij}^{\,\mathcal{B}}$
is also symmetric
(see also \cite{Kruyt:2003a}).
\par
Integrand $\nu_{i}^{\,p}t_{j}(\boldsymbol{\nu}^{\,p})$
in Eq.~(\ref{eq:Bagi2})
involves a distribution of traction along~$p$'s cut
surface $\Bpart\cap\partial\mathcal{B}$.
This distribution will usually be unknown, but we
restrict each cut surface to being a flat
plane having a unit normal $\mathbf{e}_{I^{p}}$
aligned with one of the global coordinate directions
$I^{p}$.
We specify that each particle is cut by, at most,
a single planar facet of $\partial\mathcal{B}$.
These restrictions allow an approximation of
$\overline{\sigma}_{ij}^{\,\mathcal{B}}$
and an assessment of the
bounds on any associated errors.
\par
For each peripheral particle, the nine components of its
contribution $\overline{\sigma}_{ij}^{p}$
to the sub-region's stress
$\overline{\sigma}_{ij}^{\mathcal{B}}$ fall into one of
two cases.
\subsection{Case~1, $\Sigmaijp$ and
                    $\overline{\sigma}_{ji}^{\,p}$ with
                    $i= I^{p}$ and $j\in\{1,2,3\}$}\label{sec:Case1}
\begin{figure}
\centering
\includegraphics[scale=1.00]{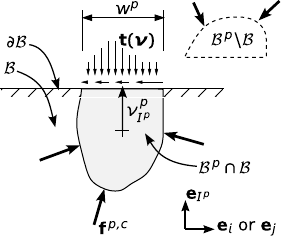}%
\caption{Calculating the stress contribution 
$\overline{\sigma}_{ij}^{\,p}$.}
\label{fig:stress-calc4b}
\end{figure}

For these five components,
the cut surface through particle $p$
has normal $\mathbf{e}_{I^{p}}$,
such that stress component
$\Sigmaijp$ is either a normal or shearing
stress on the surface.
The case also includes the complementary stress components
$\overline{\sigma}_{ji}^{\,p}$.
Along the cut surface
$\Bpart \cap \partial\mathcal{\,B}$,
the local coordinate $\nu_{I^{p}}$
of particle~$p$ in Eq.~(\ref{eq:Bagi2}b)
is constant 
(Fig.~\ref{fig:stress-calc4b}),
and, thus, offset $\nu^{p}_{I^{p}}$
is independent of traction
$t_{j}(\boldsymbol{\nu}^{p})$.
The contribution $\overline{\sigma}_{I^{p}j}^{\,p}$ is
exactly
\begin{equation} \label{eq:R}
\overline{\sigma}_{I^{p}j}^{\,p} =
\frac{1}{V^{p}}\,
\nu_{I^{p}}^{\,p}
\negthickspace\negthickspace
\underset{\Bpart \cap \partial\mathcal{B}}{\int}
\negthickspace\negthickspace
{t_{j}(\boldsymbol{\nu}^{p})\,ds}
,\quad
j\in\{1,2,3\}
\end{equation}
The integral is the total
internal force on particle~$p$ acting on the
cut surface.
Because this force is in equilibrium with the external force
$\mathbf{T}^{p}$, as in Eq.~(\ref{eq:Q2}),
the average stress on particle $p$ is exactly
%
%
%
\begin{equation} \label{eq:Case1}
\overline{\sigma}_{I^{p}j}^{\,p} =
\frac{1}{V^{p}}
\left(
-\nu_{I^{p}}^{p} T_{j}^{p}
\;+
\sum_{c\in (\partial\Bpart\cap\mathcal{B})}
\negthickspace\negthickspace
{\nu_{i}^{\,p,\,c}f_{j}^{\,p,\,c}}
\;+
\underset{\Bpart\cap\mathcal{B}}{\int}
\negthickspace\negthickspace
{\nu_{i}^{\,p}b_{j}^{\,p}\,dv}
\right)
,\quad
j\in\{1,2,3\}
\end{equation}
The complementary stresses
$\overline{\sigma}_{jI^{p}}^{\,p}$
are also exact and equal to $\overline{\sigma}_{I^{p}j}^{\,p}$.
As such, five of the nine components of 
stress $\Sigmaijp$ are exactly computed
when the cut surface is aligned with a coordinate plane.
The remaining four components are addressed in the next section.
\subsection{Case~2, $\Sigmaijp$ with
                    $i,j\neq I^{p}$}\label{sec:Case2}
As with Case~1,
we assume that the cut surface through $p$
is planar and is parallel to a
coordinate plane (with normal $\mathbf{e}_{I^{p}}$),
but we consider the contribution $\Sigmaijp$
of particle $p$ to the sub-region's stress
$\overline{\sigma}_{ij}^{\mathcal{B}}$ for the four
components having $i,j\neq I^{p}$.
Because the distribution of traction $\mathbf{t}^{p}$
is unknown along the cut surface, the traction's contribution
$\Sigmaijp$ cannot be exactly computed
for these four components
(i.e., the first integral of Eq.~\ref{eq:Bagi2}b).
\par
By centering the local coordinates $\nu_{i}^{p}$
at the centroid of the cut surface
$\Bpart\cap\partial\mathcal{B}$,
the contribution to stress $\overline{\sigma}_{ij}^{\mathcal{B}}$
of the surface's traction
(again, the first integral in Eq.~\ref{eq:Bagi2}b)
is reduced relative to the contribution of the remaining forces.
The \emph{resultant traction} along the cut surface,
$-T_{j}^{p}$,
simply counteracts
the contact and body forces $T_{j}^{p}$
that act on the remaining part
$\partial\mathcal{B}^{p}\cap\mathcal{B}$,
as in Eq.~(\ref{eq:Q2}).
The $\Sigmaijp$
can be approximated as $\widetilde{\sigma}_{ij}^{p}$
by simply ignoring the traction on the cut surface.
The same result is obtained by assuming
the traction is \emph{uniform} across this cut surface,
from which the estimated stress contribution of the traction is zero
(since $\nu^{p}_{i}$ is centered on the surface),
and the stress is exactly computed:
$\Sigmaijp=\widetilde{\sigma}_{ij}^{\,p}$.
A more reasonable estimate of the error
is based on an assumption that 
the traction along the cut surface
\emph{acts in a uniform direction} but
with a varying magnitude
(that is, by assuming
there is no reversal in the sign of $t_{j}^{p}$
across $\Bpart\cap\partial\mathcal{B}$,
such that $\int t_{j}\,ds=T_{j}^{p}$).
If this is the case,
the estimate $\widetilde{\sigma}_{ij}^{p}$
has an error bounded by
\begin{equation}\label{eq:bounds}
\left|\,\Sigmaijp - \widetilde{\sigma}_{ij}^{\,p}\right|
\le
\frac{1}{V^{p}}\,
w_{i}^{p}\,
|T_{j}^{p}|
,\quad
i,j\neq I^{p}
\end{equation}
where $w_{i}^{p}\le|\nu_{i}^{p}|$
is the (positive) width of the cut surface
in direction $\mathbf{e}_{i}$.
The inequality~(\ref{eq:bounds})
follows from
$|\int\nu_{i}^{p}t_{j}\,ds|
\le\int|\nu_{i}^{p}t_{j}|\,ds
=\int|\nu_{i}^{p}||t_{j}|\,ds
\le w_{i}^{p}\int|t_{j}^{p}|\,ds
=w_{i}^{p}|T_{j}^{p}|$.
Note, however, that
the error is unbounded
if the traction has a general distribution
with $t_{j}^{p}$ possibly changing direction across the cut surface.
\section{\normalsize Discussion}\label{sec:discussion}
With planar cuts through peripheral particles,
the stress contribution $\Sigmaijp$
\emph{of a single particle $p$}
to the sub-region's average stress
$\overline{\sigma}_{ij}^{\mathcal{B}}$ is exactly
computed for five of the nine stress components
(Case~1).
For the remaining components,
the contributions of each peripheral particle
can only be approximated (Case~2).
As a result,
all components of the aggregate, cumulative stress
$\overline{\sigma}_{ij}^{\mathcal{B}}$ 
in Eq.~(\ref{eq:Bagi2}) will be approximate if even a single
particle has a
cut surface that is aligned
with $i,j\neq I^{p}$.
This seemingly broad limitation does, however, have some fortunate
exceptions, which can be exploited
when selecting a sub-region:
\begin{itemize}
\item
If the boundary $\partial\mathcal{B}$ of the
sub-region only passes through the \emph{contact points}
between particles,
then the widths of cut surfaces
$w^{p}_{i}$ in Eq.~(\ref{eq:bounds})
are zero and an exact average stress
$\overline{\sigma}_{ij}^{\mathcal{B}}$
is computed for all nine components
with Love--Weber equation
(see references in Section~\ref{sec:introduction}).
\par\quad
This choice of a boundary $\partial\mathcal{B}$
is further relaxed when only particular components
of $\overline{\sigma}_{ij}^{\mathcal{B}}$ require exact
calculation.
In this case,
only certain cut
surfaces must avoid particle bodies and pass through the contact points: 
those cut surfaces having a normal aligned with
$\mathbf{e}_{I}$, 
such that $i,j\neq I$.
All other cut surfaces may pass through peripheral particles.
\item
When a sub-region $\mathcal{B}$ is \emph{periodic}, 
certain stress components are exactly computed.
\begin{figure}
  \centering
  \includegraphics[scale=1.00]{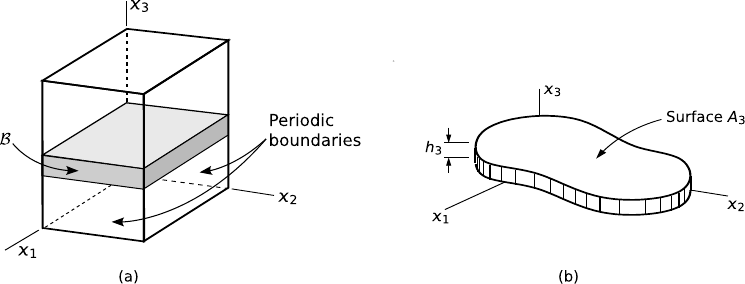}%
  \caption{Special cases with reduced error:
           (a)~sliced sub-region $\mathcal{B}$ 
           with periodic boundaries;
           (b)~thin, flat sub-region
           having $I=3$}
  \label{fig:Cases}
\end{figure}
In Fig.~\ref{fig:Cases}a, 
sub-region $\mathcal{B}$
extends across periodic boundaries in the $x_{1}$ and $x_{2}$
directions.
The five stress components 
$\overline{\sigma}_{33}^{\mathcal{B}}$,
$\overline{\sigma}_{13}^{\mathcal{B}}$,
$\overline{\sigma}_{23}^{\mathcal{B}}$,
$\overline{\sigma}_{31}^{\mathcal{B}}$, and
$\overline{\sigma}_{32}^{\mathcal{B}}$
can be computed exactly, since the periodic boundaries eliminate any
cut surfaces having normals $\mathbf{e}_{1}$ and $\mathbf{e}_{2}$.
\item
The errors of certain components
$\overline{\sigma}_{ij}^{\mathcal{B}}$ are minimized
by reducing the sub-region's width.
For example, Fig.~\ref{fig:Cases}b shows
a thin prismatic sub-region $\mathcal{B}$
between two flat and parallel boundaries of area
$A_{I}$ (here, $I=3$).
The two flat caps have an orientation $\mathbf{e}_{I}$
and are separated by thickness $h_{I}$.
The local $\nu_{I}$ coordinate can be centered on one of the two
surfaces, say the ``bottom'' surface.
As thickness $h_{I}\rightarrow 0$ is reduced,
the lateral surface cuts through fewer particles.
Indeed, one can
choose surface $A_{I}$ so that its outer rim completely
surrounds the sub-region's particles.
Five stress components, $\overline{\sigma}_{Ij}^{\mathcal{B}}$
and $\overline{\sigma}_{jI}^{\mathcal{B}}$,
correspond to Case~1,
and, in the limit of a reduced thickness $h_{I}$,
the average stress across surface $A_{I}$,
as in Eq.~(\ref{eq:Case1}), is
\begin{subequations}\label{eq:limit}
\begin{align}
\underset{h_{I}\rightarrow 0}{\text{lim}}\,
\overline{\sigma}_{Ij}^{\mathcal{B}} 
&=
\underset{h_{I}\rightarrow 0}{\text{lim}}
\frac{1}{A_{I} h_{I}}
\left(
-\underset{p\in\,\text{top}}{\sum}
h_{I} T_{j}^{p}
+
\sum_{c\in(\partial\Bpart\cap\mathcal{B})}
\negthickspace\negthickspace
{\nu_{i}^{p,\,c}f_{j}^{p,\,c}}
+
\underset{\Bpart\cap\mathcal{B}}{\iint}
\negthickspace
{\nu_{I}^{p} b_{j}^{p} \,d\nu_{I}\,ds}
\right)
\\
&=
-\frac{1}{A_{I}}
\underset{p\in\,\text{top}}{\sum}
T_{j}^{p} \;, 
\end{align}
\end{subequations}
where the summation 
includes only
particles that are cut by the top surface
($p\in\text{top}$,
with the bottom surface eliminated, since its
$\nu_{I}$ coordinate is zero).
%
%
The second term on the right of Eq.~(\ref{eq:limit}a)
approaches zero, as the sub-region includes fewer contacts;
the final term in Eq.~(\ref{eq:limit}a) also approaches zero,
since $0<\nu_{I}<h_{I}$.
The result in Eq.~(\ref{eq:limit}b) is not surprising:
by minimizing thickness $h_{I}$,
the average traction
$\overline{\sigma}_{Ij}^{\mathcal{B}}$ along the planar surface
equals the sum of the forces acting on the particles along one side,
divided by the surface's area.
\end{itemize}
\par
In general,
the error in estimating a component
$\overline{\sigma}_{ij}^{\,\mathcal{B}}$ 
results from inexact
approximations $\widetilde{\sigma}_{ij}^{\,p}$
of contributions from particles
having cut surface's
that are aligned with neither
the $\mathbf{e}_{i}$ nor $\mathbf{e}_{j}$ directions
(Case~2, Section~\ref{sec:Case2}).
An error in $\overline{\sigma}_{ij}^{\,\mathcal{B}}$ ensues
when even a single particle $p$ manifests this condition.
We gain an estimate of the
relative error that is introduced by all such cut surfaces,
by adopting the reasonable assumption that
tractions along the particles'
cut surfaces act in uniform directions
(as with Eq.~\ref{eq:bounds}).
If so, the total error of estimating a
sub-region's stress $\overline{\sigma}_{ij}^{\,\mathcal{B}}$
by using approximations of the
particles' stresses, $\widetilde{\sigma}_{ij}^{\,p}$,
is bounded by a sum of the particle errors:
%
\begin{equation} \label{eq:deltaSigma}
|\overline{\sigma}_{ij}^{\,\mathcal{B}}
 - \widetilde{\sigma}_{ij}^{\,\mathcal{B}} | \leq \;
\frac{1}{V^{\mathcal{B}}}
\underset{\substack{p\,:\: I^{p}\neq i,j \\ \mathcal{B}^{p}\cap\partial\mathcal{B}\neq\varnothing}}{\sum}
\!\!\!
w_{i}^{p}\, |T_{j}^{p}|
\; .
\end{equation}
where the sum includes only peripheral particles
(i.e., $\mathcal{B}^{p}\cap\partial\mathcal{B}\neq\varnothing$)
with cut surfaces of orientation $\mathbf{e}_{I^{p}}$,
$i,j\neq I^{p}$.
If we neglect body forces $b_{j}^{p}$ in $T_{j}^{p}$ and
assume that
the average magnitude of contact forces
$|f^{p,c}_{j}|$ is roughly
the same for interior and peripheral particles,
and if we also assume that the cut surface widths
$w_{i}^{p}$ are less than
the particle sizes themselves (and the distances
$|\nu_{i}^{p,c}|$ in Eq.~\ref{eq:Bagi2}),
the relative error 
is the quotient of Eqs.~(\ref{eq:deltaSigma})
and~(\ref{eq:Bagi2}), bounded by
\begin{equation} \label{eq:W}
\left| \frac{\overline{\sigma}_{ij}^{\,\mathcal{B}}
 - \widetilde{\sigma}_{ij}^{\,\mathcal{B}}}
{\widetilde{\sigma}_{ij}^{\,\mathcal{B}}} \right|
\;\lesssim\;
\frac{M_{\text{ext}, I\neq i,j}}
{M_{\text{int}}}\;,
\end{equation}
which is the ratio of the number of contacts
$M_{\text{ext}, I\neq i,j}$
on the external portions of peripheral particles
(contacts on the dashed portions
$\mathcal{B}^{p}\backslash\mathcal{B}$
in Fig.~\ref{fig:stress-calc4b}) and the number of
contacts $M_{\text{int}}$ inside of $\mathcal{B}$.
The first count 
only includes cut surfaces whose orientation $\mathbf{e}_{I}$
is aligned in neither of the coordinate directions $i$ or $j$.
\par
The relative error in Eq.~(\ref{eq:W}) can, of course,
be reduced by
increasing the size of the sub-region $\mathcal{B}$, 
since the ratio in Eq.~(\ref{eq:W})
is roughly proportional to 
the inverse of the sub-region's width in the direction $I\neq i,j$.
That is, the error in estimating
$\overline{\sigma}^{\mathcal{B}}_{ij}$
can be reduced by choosing sub-regions
with a minimal width in direction $\mathbf{e}_{I}, I\neq i,j$,
a choice taken to its limit in the example of
Fig.~\ref{fig:Cases}b and~Eq.~(\ref{eq:limit}).
\section{\normalsize Conclusions}\label{sec:conclusions}
Stress is a continuum quantity that applies to material points,
and we demonstrate one shortcoming of extending its definition
to material sub-regions:
the volume-averaged stress cannot, \emph{in general}, be
computed exactly from discrete force quantities, such as
the contact and body forces.
This paper, however,
advises conditions in which a sub-region may be chosen
to reduce the error or, with careful selection of
the sub-region, eliminate error
altogether.
%
\section*{\normalsize Acknowledgment}
The author wishes to thank Dr. K. Bagi for reviewing
an early draft of this paper.
%
%

\begin{thebibliography}{10}
\expandafter\ifx\csname url\endcsname\relax
  \def\url#1{\texttt{#1}}\fi
\expandafter\ifx\csname urlprefix\endcsname\relax\def\urlprefix{URL }\fi
\expandafter\ifx\csname href\endcsname\relax
  \def\href#1#2{#2} \def\path#1{#1}\fi

\bibitem{Love:1927a}
A.~E.~H. Love, A Treatise on the Mathematical Theory of Elasticity, 4th
  Edition, Dover Pub., New York, N.Y., 1927.

\bibitem{Weber:1966a}
J.~Weber, Recherches concernant les contraintes intergranulaires dans les
  milieux pulv{\'e}rulents, Bulletin de Liaison des Ponts-et-chauss{\'e}es 20
  (1966) 1--20.

\bibitem{Rothenburg:1981a}
L.~Rothenburg, A.~P.~S. Selvadurai, A micromechanical definition of the
  {C}auchy stress tensor for particulate media, in: A.~P.~S. Selvadurai (Ed.),
  Mechanics of Structured Media, Part B, Elsevier, Amsterdam, The Netherlands,
  1981, pp. 469--486.

\bibitem{Bagi:1996a}
K.~Bagi, Stress and strain in granular assemblies, Mech. of Mater. 22~(3)
  (1996) 165--177.
\newblock \href
  {http://dx.doi.org/https://doi.org/10.1016/0167-6636(95)00044-5}
  {\path{doi:https://doi.org/10.1016/0167-6636(95)00044-5}}.

\bibitem{Satake:2004a}
M.~Satake, Tensorial form definitions of discrete-mechanical quantities for
  granular assemblies, Int. J. Solids Struct. 41~(21) (2004) 5775--5791.
\newblock \href
  {http://dx.doi.org/https://doi.org/10.1016/j.ijsolstr.2004.05.046}
  {\path{doi:https://doi.org/10.1016/j.ijsolstr.2004.05.046}}.

\bibitem{Bardet:2001a}
J.~P. Bardet, I.~Vardoulakis, The asymmetry of stress in granular media, Int.
  J. Solids Struct. 38~(2) (2001) 353--367.
\newblock \href
  {http://dx.doi.org/https://doi.org/10.1016/S0020-7683(00)00021-4}
  {\path{doi:https://doi.org/10.1016/S0020-7683(00)00021-4}}.

\bibitem{Chang:2005a}
C.~S. Chang, M.~R. Kuhn, On virtual work and stress in granular media, Int. J.
  Solids Struct. 42~(13) (2006) 6026--6051.
\newblock \href
  {http://dx.doi.org/https://doi.org/10.1016/j.ijsolstr.2004.11.011}
  {\path{doi:https://doi.org/10.1016/j.ijsolstr.2004.11.011}}.

\bibitem{Kruyt:2003a}
N.~P. Kruyt, Statics and kinematics of discrete {C}osserat-type granular
  materials, Int. J. Solids Struct. 40~(3) (2003) 511--534.
\newblock \href
  {http://dx.doi.org/https://doi.org/10.1016/S0020-7683(02)00624-8}
  {\path{doi:https://doi.org/10.1016/S0020-7683(02)00624-8}}.

\bibitem{Froiio:2006a}
F.~Froiio, G.~Tomassetti, I.~Vardoulakis, Mechanics of granular materials: The
  discrete and the continuum descriptions juxtaposed, Int. J. Solids Struct.
  43~(25--26) (2006) 7684--7720.
\newblock \href
  {http://dx.doi.org/https://doi.org/10.1016/j.ijsolstr.2006.03.023}
  {\path{doi:https://doi.org/10.1016/j.ijsolstr.2006.03.023}}.

\bibitem{Bagi:1999a}
K.~Bagi, Microstructural stress tensor of granular assemblies with volume
  forces, J. Appl. Mech. 66~(4) (1999) 934--936.
\newblock \href {http://dx.doi.org/https://doi.org/10.1115/1.2791800}
  {\path{doi:https://doi.org/10.1115/1.2791800}}.

\bibitem{Nicot:2013b}
F.~Nicot, N.~Hadda, M.~Guessasma, J.~Fortin, O.~Millet, On the definition of
  the stress tensor in granular media, Int. J. Solids Struct. 50~(14-15) (2013)
  2508--2517.
\newblock \href
  {http://dx.doi.org/https://doi.org/10.1016/j.ijsolstr.2013.04.001}
  {\path{doi:https://doi.org/10.1016/j.ijsolstr.2013.04.001}}.

\bibitem{Yan:2019a}
B.~Yan, R.~A. Regueiro, Definition and symmetry of averaged stress tensor in
  granular media and its {3D} {DEM} inspection under static and dynamic
  conditions, Int. J. Solids Struct. 161 (2019) 243--266.
\newblock \href
  {http://dx.doi.org/https://doi.org/10.1016/j.ijsolstr.2018.11.021}
  {\path{doi:https://doi.org/10.1016/j.ijsolstr.2018.11.021}}.

\bibitem{Gaspar:2001a}
N.~Gaspar, M.~A. Koenders, Micromechanic formulation of macroscopic structures
  in a granular medium, J. Eng. Mech. 127~(10) (2001) 987--993.
\newblock \href
  {http://dx.doi.org/https://doi.org/10.1061/(ASCE)0733-9399(2001)127:10(987)}
  {\path{doi:https://doi.org/10.1061/(ASCE)0733-9399(2001)127:10(987)}}.

\bibitem{Gaspar:2002a}
N.~Gaspar, Structures and heterogeneity in deforming, densely packed granular
  materials, Ph.{D}. thesis, Kingston University, Kingston, U.K. (2002).

\bibitem{Kuhn:2002a}
M.~R. Kuhn, Are granular materials simple? {A}n experimental study of strain
  gradient effects and localization., Mech. of Mater. 37~(5) (2005) 607--627.
\newblock \href
  {http://dx.doi.org/https://doi.org/10.1016/j.mechmat.2004.05.001}
  {\path{doi:https://doi.org/10.1016/j.mechmat.2004.05.001}}.

\end{thebibliography}
%
%

\end{document}